\documentclass[pre,twocolumn,notitlepage,longbibliography,amsmath,amssymb,floats,superscriptaddress,nofootinbib,10pt]{revtex4-1}
\usepackage{graphicx,color}
\usepackage{amsfonts, amsbsy}
\usepackage{amssymb}
\usepackage{amsmath}
\usepackage{eqnarray}
\usepackage{bm}
\usepackage{blkarray}
\usepackage{mathtools}
\usepackage{psfrag}
\usepackage{caption}
\usepackage{subcaption}
\usepackage{multirow}
\usepackage{textcomp}
\usepackage{units}
\usepackage{lipsum}
\usepackage[title]{appendix}
\usepackage{soul}
\usepackage{titlesec}
\usepackage{times}
\usepackage{enumitem}
\usepackage[breaklinks,colorlinks = true,linkcolor = blue,urlcolor  = blue,citecolor = blue,anchorcolor = blue]{hyperref}
\usepackage{float}

\begin{document}
\title{Elastic Li\'{e}nard-Wiechert potentials of dynamical dislocations from tensor gauge theory in 2+1 dimensions}
\author{Lazaros Tsaloukidis}
\email{ltsalouk@pks.mpg.de}
\affiliation{Max Planck Institute for the Physics of Complex Systems, N\"othnitzer Stra{\ss}e 38, 01187 Dresden, Germany}
\affiliation{W\"urzburg-Dresden Cluster of Excellence ct.qmat, 01187 Dresden, Germany}
\author{Piotr Sur\'{o}wka}
\email{piotr.surowka@pwr.edu.pl}
\affiliation{Institute of Theoretical Physics, Wroc\l{}aw University of Science and Technology, 50-370, Wroc\l{}aw, Poland}
\date{\today}

\begin{abstract}
The dynamics of defect excitations in crystalline solids is necessary to understand the macroscopic low-energy properties of elastic media. We use fracton-elasticity duality to systematically study the defect dynamics and interactions in the linear isotropic medium. We derive the explicit expressions for the dual gauge potentials for moving dislocations and the resulting Jefimenko equations. We also compute stresses and strains. The paper includes two physical situations: when the vacancy number is fixed and when the number is fluctuating. If defects are present we show a constraint that needs to be satisfied by them when they climb perpendicularly to their Burgers vector. Next, we extend the classic result of Peach and Koehler for the force between two dislocations and show that, similarly, to moving charges in electrodynamics, it is non-reciprocal, when one dislocation is moving. We argue that our formalism can be extended beyond Cauchy's elasticity by exploiting the simplifications provided by the dual gauge formulation of elastic stresses. 
\end{abstract}

\maketitle

\section{Introduction}

Stresses induced by dislocations play an important role in many fundamental physical phenomena in crystals and thin crystalline films. This role of topological defects is manifested for example in plastic, electronic or optical characteristics. In order to have a good understanding of phenomena involving dislocations one needs to shed light on their dynamics. This is a classic problem in the theory of elasticity that was initiated in its early days and subsequently explored by many works (see e.g. \cite{eshelby_uniformly_1949,kiusalaas_elastic_1964,weertman_stress_1967,brock_dynamic_1982,markenscoff_nonuniformly_1981,markenscoff_accelerating_1982,markenscoff_dislocation_1983,brock_non-uniform_1983,pellegrini_dynamic_2010,lazar_elastic_2011,lazar_nearfield,lazar_elastodynamic_2012,lazar_retardation_2013,lazar_distributional_2016, hirth_theory_1992,kazinski_self-interaction_2022}). In order to fully solve this problem one must account for both external stresses, and self-stresses due to the movement of a dislocation. This is a demanding task in the usual formulation of elasticity. 

In this paper the aim is to address this problem using the so-called fracton-elasticity dualities, which depart from displacement formulation of elasticity (for recent reviews see \cite{grosvenor_space-dependent_2022,Gromov:2022cxa}). The dualities originate from the works on two-dimensional crystals for which the low-energy elastic degrees of freedom can be written down as gauge variables with two-indices \cite{kleinert_duality_1982,kleinert_double_1983,beekman_dual_2017-1,beekman_dual_2017}. More recently this original framework of elastic dualities was refined, and topological defects, in the absence of crystal vacancies, were interpreted as fractons, i.e. excitations with mobility constraints \cite{pretko_fracton-elasticity_2018,pretko_crystal--fracton_2019,radzihovsky_fractons_2020,radzihovsky_quantum_2020,gromov_duality_2020,surowka_dual_2021,hirono_effective_2022,caddeo_emergent_2022,zaanen_crystal_2022}. Such excitations act as sources for the tensor gauge fields. The most important part of this construction are convenient low-energy degrees of freedom together with symmetry principles that precisely capture the dynamics of defects and, allow one to get insight based purely on symmetry considerations \cite{pretko_the_2018,gromov_towards_2019}. A successful application of this line of reasoning has been in the quest of constructing hydrodynamic theories of kinematically constrained constituents \cite{gromov_fracton_2020,grosvenor_hydrodynamics_2021,Glorioso_breakdown_22,burchards_coupled_2022,Glodkowski:2022xje,Glorioso:2023chm}. 

The traditional approach to two-dimensional elasticity is based on plane-strain elasticity, which assumes that the strain in one direction is zero. The dualities reformulate elasticity in the language of gauge theories. The gauge fields then serve to encapsulate the stress states experienced within two-dimensional thin films, drawing parallels to the formulation of plane stress in these structures. The concept of plane stress is applicable when there is no stress exerted along a particular direction - typically the thickness. In essence, the material remains unaffected by any external forces acting perpendicular to its plane. This results in the complete absence of both the direct stress in that specific direction, as well as any shear stresses that would otherwise emerge due to interaction with that direction.  In the simplest incarnation a moving topological charge in a stress gauge theory is a generalization of the moving electromagnetic charge in U(1) gauge theories. This problem was solved by Li\'{e}nard and Wiechert who computed electromagnetic potentials of an arbitrarily moving relativistic electric charge (see e.g. \cite{zangwill_modern_2013}) \footnote{A similar problem was also addressed in the context of moving vortices without any reference to gauge fields \cite{radzihovsky_anomalous_2015}}. Later Jefimenko generalized these results to give electric and magnetic fields due to a distribution of electric charges and electric currents in space \cite{jefimenko_electricity_1989}. Therefore, a dual gauge theory formulation of elasticity offers a unique opportunity to understand the dynamics and stresses generated by a distribution of elastic defects. As we will see, separate solutions can be obtained in materials depending on whether the number of vacancies is fixed or not. In the former case the dislocations are kinematically restricted to move along their Burgers vector. In the latter case the dislocations are not constrained, however, their movement perpendicular to the Burgers vector is tightly connected to the distribution of vacancies in the material. Note that in the context of strain formulation of elasticity, the elastodynamic Li\'{e}nard–Wiechert potentials and elastic fields of non-uniformly moving point and line forces were also derived in analogy to the electromagnetic Li\'{e}nard–Wiechert potentials \cite{lazar_elastodynamic_2012}.

Modern studies of defect dynamics go much beyond original crystalline systems. Examples include topological defects in rheology \cite{furst_microrheology_2017} or active matter \cite{bowick_symmetry_2022,shankar_topological_2022}. In rheology viscoelastic properties for incompressible systems can be derived by the correspondence principle upon changing the real elastic parameters to complex transport coefficients \cite{pipkin_lectures_1986}. Therefore, a good understanding of elasticity is a convenient starting point to understand more complex viscoelastic materials.  Finally defects in active matter naturally emerge in fluids that form from collections of bacterial suspensions \cite{sokolov_emergence_2019} or other macroscopics ensembles of living organisms \cite{duclos_topological_2017,tan_odd_2022}. More recently topological defects appeared in the context of active solids and metamaterials \cite{braverman_topological_2021}.

In this paper we show that the dynamics of topological defects in isotropic elasticity on a plane can be simplified considerably if elasticity is formulated as a gauge theory. Exploiting this simplification, we derive the stresses that follow from the dynamics of arbitrarily moving dislocation. We also include time dependence into the classic result of the Peach-Koehler force that captures the interactions and external stresses acting on a dislocation \cite{peach_forces_1950}. In consequence our formalism can be generalized to study more complex theories of elasticity or viscoelasticity. 

Our paper is organized as follows. In Sec. \ref{Sec2} we review the fracton-elasticity duality. In Sec. \ref{Sec3} we derive equations of motion for the dual gauge potentials. Sec. \ref{Sec4} introduces the notion of defects. In Sec. \ref{Sec5} we present the solutions and discuss physical implications of the defect motion. Sec. \ref{Sec6} is devoted to interactions between dislocations. We derive the expression for the dynamical Peach-Koehler force between two dislocations. Finally in Sec. \ref{Sec7} we close with conclusions and discussion.
 
\section{Fracton-Elasticity Duality}\label{Sec2}

We start the theoretical study with the two-dimensional low-energy elasticity theory action:
\begin{equation}
S=\frac{1}{2}\int\left[\rho_d(\partial_t u_i)^2-C_{ijkl}u_{ij}u_{kl}\right]d^2xdt,
\end{equation}
where $u_{ij}=\frac{1}{2}(\partial_i u_j+\partial_j u_i)$ is the linear part of the symmetric strain tensor, with $u_j$ the displacement field and $C_{ijkl}$ the first-gradient elasticity matrix, with its components determined by the type of medium at hand. We use Einstein's summation convention, where $i,j,k,l=1,2$. In order to capture the defects in this theory we are interested in the singular part $u_{ij}^{s}$ of the strain tensor field $u_{ij}=u_{ij}^{s}+\frac{1}{2}(\partial_i \overline{u}_j+\partial_i \overline{u}_j)$, with the smooth part later integrated out providing a constraint. 
Generally speaking for a given set of conjugate fields $\psi $ and $\phi$, we can compactly write the functional 
\begin{equation*}
Z=\int \mathcal{D}\psi e^{-S},
\end{equation*}
of the theory by performing a Hubbard-Stratonovich transformation, as 
\begin{equation*}
\exp \left[\frac{1}{2} {\psi M \psi} \right]=\frac{1}{\mathcal{N}}   \int D \phi \exp \left[{\frac{-1}{2}  \phi M^{-1} \phi + \psi \phi} \right], \,
\end{equation*}
 where $\mathcal{N}$ is a normalization factor. This way we receive the action in the dual variables \cite{kleinert_gauge_1989}:
\begin{equation}
S=\int \left[\frac{1}{2}C^{-1}_{ijkl}\sigma^{ij}\sigma^{kl}-\frac{1}{2\rho_d}\pi^i\pi_i-\sigma^{ij}u_{ij}+\pi^i\partial_tu_i\right]d^2xdt.
\end{equation}
where $\sigma_{ij}$ is the elasticity stress tensor with the symmetry $\sigma_{ij}=\sigma_{ji}$, and $\pi_i$ is the elastic momentum vector. We employ Einstein's summation convention. The details of inversion of the elastic tensor are presented in Appendix \ref{AppA}. If the smooth and singular parts of the symmetric strain tensor are now plugged into the equation above, the smooth part is then integrated out, giving the well known Hooke law:
\begin{equation}
\partial_t\pi^i-\partial_j\sigma^{ij}=0.
\end{equation}
We now introduce the two rotated fields:
\begin{equation}
B^i=\epsilon^{ij}\pi_j \qquad E^{ij}_{\sigma}=\epsilon^{ik}\epsilon^{jl}\sigma_{kl} .
\label{rotated}
\end{equation}
which along with the two-dimensional Levi-Civita identity
$\epsilon^{ij}\epsilon_{jk}=\delta^i_k$ transform the above equation into Faraday Law:
\begin{equation}
\partial_t B_i+\epsilon_{jk}\partial^jE^{ki}_{\sigma}=0.
\end{equation}
It must be stated that the lower case index $\sigma$ on the rank-2 tensor electric field, signifies the relation with the rotated stress field. This field is actually the elastic counterpart of the usual electric field used in EM theory and works the same way as the electric displacement field. The relation connecting one another is given through the formula
\begin{equation}
E^{ij}=\tilde {C}_{ijkl}^{-1}E^{kl}_{\sigma},
\end{equation}
where $\tilde {C}_{ijkl}^{-1}=\epsilon_{im}\epsilon_{jn}\epsilon_{ko}\epsilon_{lp}C_{mnop}^{-1}$
is the rotated inverted elasticity tensor.\\
Faraday's equation can be solved by introducing the generalization of the EM theory into a rank-2 gauge theory through:
\begin{equation}
B^i=\epsilon_{jk}\partial^jA^{ki} ,\qquad  E^{ij}_{\sigma}=-\partial_t A^{ij}-\partial_i\partial_j\phi,
\label{gaugefields}
\end{equation}
where  $\phi$ is the usual scalar fractonic field, and $A_{ij}$ is a symmetric tensor gauge field. Of course Faraday's equation stays invariant under the gauge transformation:
\begin{equation}
A_{ij}\rightarrow A_{ij}+\partial_i\partial_jf(x_k,t), \qquad \phi \rightarrow\phi-\partial_t f(x_k,t),
\end{equation}
where $f(x_k,t)$ is an arbitrary space-time function. Using the gauge fields now, we can rewrite the action of our theory, to also include currents and sources
\begin{equation}
S=\int\left[\frac{1}{2}\tilde{C}_{ijkl}^{-1}E^{ij}_{\sigma}E^{kl}_{\sigma}-\frac{1}{2\rho_d}B^iB_i-\rho\phi+J^{ij}A_{ij}\right]d^2xdt.
\label{FE-ACTION-1}
\end{equation}
The above equation describes isolated charges of the theory in the form of disclinations. Since our main concerns are dislocation defects, we introduce the vector fractonic potential $\phi_i=\partial_i \phi$ and the dislocation charge density $\rho=-\partial_i\rho^i$. By performing integration by parts on \eqref{FE-ACTION-1} we get the required action describing gauge fields sourced by dislocations
\begin{equation}
S=\int\left[\frac{1}{2}\tilde{C}_{ijkl}^{-1}E^{ij}_{\sigma}E^{kl}_{\sigma}-\frac{1}{2\rho_d}B^iB_i-\rho^i\phi_i+J^{ij}A_{ij}\right]d^2xdt.
\label{FE-ACTION-2}   
\end{equation}
We have successfully managed to map the low-energy action of the elasticity theory containing two gapless phonon modes (longitudinal and transverse motion) into another U(1) vector charge gauge theory containing two gapless gauge modes.
With the usage of the first (Gauss) and the fourth (Amp\`ere) Maxwell law for the duality, we can extract the continuity equation for the vector charge theory
\begin{equation}
\partial_t \rho_j+\partial_iJ_{ij}=0.
\end{equation}
\section{Equations of motions for isotropic materials in two dimensions}\label{Sec3}
We start by writing the Lagrangian for elasticity in the language of gauge fields by choosing to have $\rho_d=1$, following the convention of Pretko:
\begin{equation}
\mathcal{L}=\frac{1}{2}(\tilde{C}_{ijkl}^{-1}E^{ij}_{\sigma}E^{kl}_{\sigma}-B^iB_i)-\rho^i\phi_i+J^{ij}A_{ij},
\label{Lagrangian}
\end{equation}
where for the case of an isotropic material we have the formula for the inverse elasticity tensor as a function of the elastic bulk modulus $\kappa$ and the shear modulus $\mu$,
\begin{equation}
\tilde{C}_{ijkl}^{-1}=\left(\frac{1}{4\kappa}-\frac{1}{4\mu}\right)\delta_{ij}\delta_{kl}+\frac{1}{4\mu}(\delta_{ik}\delta_{jl}+\delta_{il}\delta_{jk}).
\label{elasticitytensor}
\end{equation}
The above tensor obeys the Maxwell-Betti reciprocity relations
$\tilde{C}_{ijkl}^{-1}=\tilde{C}_{jikl}^{-1}=\tilde{C}_{ijlk}^{-1}=\tilde{C}_{jilk}^{-1}$. By plugging Eq. \eqref{gaugefields} into Eq. \eqref{Lagrangian} and using the Euler-Lagrange equations, we extract the two equations of motion (Gauss law and Amp\`ere law)
\begin{equation}
2C_2\partial_t\partial_i A_{ij}+C_2\partial_i^2\phi_j+C_1\partial_t\partial_j A_{kk}+C_3\partial_j\partial_k\phi_k+\rho_j=0,
\label{phifield}
\end{equation}
\begin{equation*}
2C_2\partial_t^2A_{ij}+2C_2\partial_t\partial_i\phi_j+C_1\big[\partial_t^2A_{kk}+\partial_t\partial_k\phi_k\big]\delta_{ij}-\partial_k^2 A_{ij}+    
\end{equation*}
\begin{equation}
\hspace{4.5cm}+\partial_i\partial_k A_{kj}-J_{ij}=0,
\label{alphafield}
\end{equation}
where the coefficients depend on the elastic moduli in the following way
\begin{equation*}
C_1=\frac{(\mu-\kappa)}{4\mu\kappa}, \qquad C_2=\frac{1}{4\mu}, \qquad C_3=\frac{1}{4\kappa}.
\end{equation*}
Both Eq.\eqref{phifield} and \eqref{alphafield} are coupled, in analogy to classical electrodynamics. In addition the trace of the tensor gauge field $A_{ij}$ is present, signifying its relation to the motion of dislocations as we show later on.

In order to simplify the equations we need to fix a gauge. We note that in the analogous problem in electrodynamics it is convenient to choose the Lorenz gauge in order to decouple the equations. Therefore we intend to generalize the usual Lorenz gauge to the case of tensor gauge theories. This can be done as follows
\begin{equation}
\partial_i A_{ij}+\frac{1}{2\mu}\partial_t\phi_j=0.
\label{Lorenz}
\end{equation}
With this choice the second term in \eqref{alphafield} cancels with the contribution of the fifth term.
In all the equations above, the density is equal to unity. Otherwise in the gauge condition \eqref{Lorenz}, the second term shall be multiplied by $\rho_d$.

The next step in deriving the wave equations uses the fact that the vector potential $\phi_i$ is given by the gradient of the scalar field $\phi$, meaning that permutation of the indices in the derivatives is allowed
$\partial_j(\partial_i\phi_i)= \partial_i^2\phi_j.$
The final forms of the wave equations for the vector field $\phi_i$ and the tensor gauge field $A_{ij}$ that contain the trace read
\begin{equation}
\Box \phi_j+\left(\frac{\mu-\kappa}{\mu+\kappa}\right)\partial_t\partial_j A_{kk}+\left(\frac{4\mu\kappa}{\mu+\kappa}\right)\rho_j=0,
\label{wave-phi}
\end{equation}
\begin{equation}
\Box A_{ij}-\left(\frac{\mu-\kappa}{4\kappa\mu}\right)\left[\partial_t^2A_{kk}+\partial_t\partial_k\phi_k\right]\delta_{ij}+J_{ij}=0,
\label{wave-alpha}
\end{equation}
where $\Box=\nabla^2-(1/\upsilon^2)\partial_t^2$, with the velocity for the $\phi_j$ field being equal to $\upsilon_{\phi}=\sqrt{\frac{\mu(\kappa+\mu)}{\kappa}}$ and with that for the $A_{ij}$ field equal to $\upsilon_A=\sqrt{2\mu}$. The result is not surprising but a few clarifications need to be made. These velocities are the dual of the ones that appear in the regular elasticity theory of a two-dimensional isotropic solid. The phonon fields there propagate with velocities equal to $\upsilon_L=\sqrt{\frac{\kappa+\mu}{\rho_d}}$ and $\upsilon_T=\sqrt{\frac{\mu}{\rho_d}}$ for a longitudinal and transverse wave propagation respectively. The fracton vector charge density is now multiplied with a factor containing the elastic constants although that is not true for the currents of our theory. We emphasize that the above differential equations do not have the desired form of wave equations yet as there appear to be contributions coming from the trace of $E^{ij}_{\sigma}$. These can be removed either by constraining the elastic coefficients $\mu=\kappa$, or by fixing a gauge that removes these contributions. In our paper we choose the latter approach by imposing the tracelessness condition, which is tied to the singular behavior of strains, as will be further elucidated in Section \ref{Sec5}. In the next section, we provide a concise overview of how defects are integrated into the theory of elasticity.

\section{Defects}\label{Sec4}
Topological defects such as disclinations or dislocations, act as sources for the singular part of strain tensor $u_{ij}$. The disclination density is given by
\begin{equation}
s=\epsilon_{ij}\partial_i\partial_j\theta=\epsilon_{ij}\epsilon_{kl}\partial_i\partial_j\partial_k u_l=\rho,
\end{equation}
where $\theta=\epsilon_{kl}\partial_ku_l$ is the bond angle. Disclinations represent the isolated fracton charges of our theory that cannot move. However, two opposite disclinations of the same magnitude would formulate the dipole that can move albeit with mobility restrictions dubbed dislocations. Their density is represented as a function of a particular lattice vector, called the Burgers vector
\begin{equation}
b_j=\oint_{\gamma}\partial_i u_j dr_i.
\end{equation}
This lattice vector is always perpendicular to the vector joining the two disclination defects. The line element $dr_i$ encircles a specific area on the solid that encloses the defect, forming the Burgers circuit $\gamma$.

In general a system containing monopoles, dipoles and even higher combinations of particles can be represented by a charge density of the form:
\begin{equation}
\rho=q\delta(\mathbf{r}-\mathbf{r'})-\mathbf{p}\cdot\mathbf{\nabla}\delta(\mathbf{r}-\mathbf{r'})+...  
\end{equation}
Since we are interested in the motion of a defect in the medium, we focus on a single dislocation, dropping both higher order terms and the immobile monopole. By comparing the above density with the vector density in our Lagrangian we can write
\begin{equation}
\rho_i(\mathbf{r'},t)=\epsilon_{ij}b_j\delta(\mathbf{r'}-\mathbf{r}_s(t')),
\label{fracton-density}
\end{equation}
where $\epsilon_{ij}b_j=p_i$ is the dipole moment  \footnote{In elasticity a continuum dislocation theory is considered. For a discussion of the dislocation density vector in two dimensions we refer the reader to \cite{lazar_dislocation_2013}.}. It is set to be equal to the Burgers vector but pointing perpendicular to it. This gives a current of the form
\begin{equation}
J_{ij}(\mathbf{r'},t)=\epsilon_{\left(ik\right.}\upsilon_{\left.j\right)}b_k\delta(\mathbf{r'}-\mathbf{r}_s(t')),
\label{fracton-gaugedensity}
\end{equation}
where $\upsilon_i$ is the dislocation velocity and $\mathbf{r}_s(t')$ is the dislocations trajectory as seen by the observer at retarded time $t'$. The trace of the current $J_{ii}$ is related to dislocation climb, i.e. dislocation moving perpendicular to its Burgers vector and not parallel, and is equal to zero in the case where there are no vacancies in the crystal. We discuss the static solutions of the field equations corresponding to dislocations and disclinations in Appendix \ref{AppB}. The dynamical solutions will be studied in the next section.
\section{Li\'{e}nard - Wiechert potentials and Jefimenko equations}\label{Sec5}
\subsection{Fixed number of vacancies}
We start with the simplest example dubbed traceless gauge theory \cite{pretko_generalized_2017}. We note that imposing the condition can be done following different conventions \cite{beekman_dual_2017}. We posit that the trace of the electric tensor field $E_{ii}$ equates to a constant, leading to the removal of the intermediate terms in \eqref{wave-phi} and \eqref{wave-alpha}, which contain emergent contributions from it.
From a comprehensive examination of the equations, it becomes apparent that the velocities of the two wave fields coincide. Before delving into the discussion of solutions, we assign a physical interpretation to this constraint. According to Pretko's theory, the scenario involving a vector charge is steered by the dual conservation laws of charge and dipole moment, which is readily observable through the first Gauss law.
Dislocations within this theory have been empirically shown to move solely along their Burgers vector (longitudinal motion), but not transversely. This phenomenon is attributed to the presence of vacancies/interstitials within the solid (see e.g. \cite{kumar_symmetry-enforced_2019}). When analyzing the quadrupole moment, the trace is represented as follows:
\begin{equation*}
D_{ii}=-\int_Vd^2x(\rho_i x_i+2E_{ii})=\text{const.}
\end{equation*}
Based on duality relations, we derive 
\begin{equation*}
E^{ii}=(1/2\kappa)E_{\sigma}^{ii}=\partial_iu_i=n_d,
\end{equation*}
with the smooth part taken to be very small. In this context, $n_d$ symbolizes the difference between vacancies and interstitial defects. If the total number of vacancies and interstitial, remains constant in time, then so does the trace of the quadrupole moment.
On the other hand, this conservation indicates that dipoles are restricted to movement only in directions perpendicular to their dipole moment. Moving in parallel with the dipole moment would alter $n_d$, as this movement necessitates the introduction of either an interstitial or a vacancy. This aligns with a widely recognized principle in elasticity theory, which states that dislocations are confined to moving along their Burgers vector.
As far as the gauge fields are concerned, this denotes the presence of an additional scalar condition, stated as: $-\partial_tA_{ii}-\partial_i\phi_i=2\kappa n_d$.

In the case of $n_d=0$, the equations we have turn out to be the regular two-dimensional wave equations for the two potentials
\begin{equation}\label{eq:wave1}
 \Box \phi_j+2\mu\rho_j=0, \qquad \upsilon_{\phi}=\sqrt{2\mu},
\end{equation}
\begin{equation} \label{eq:wave2}
\Box A_{ij}+J_{ij}=0, \qquad \upsilon_{A}=\sqrt{2\mu}.
\end{equation}
The problem at hand parallels the one of finding Li\'{e}nard-Wiechert potentials in three space dimensions; however, since we are on the plane the exact form of the solution differs (see \cite{boito_maxwells_2020,watanabe_integral_2015} for an analogous problem in electrodynamics). The Green's function related to the above differential equations is given as the Heaviside step function with a relativistic factor included, giving rise to a phenomenon called "afterglow" \cite{barton_elements_1989,dai_origin_2013}.

The afterglow effect is a direct consequence of the Huygens principle not being valid in 2+1 dimensions and generically in space-times in an odd number of dimensions. This can be seen from the Green's function describing the equations of motion. In 3+1 dimensions the function is proportional to  Dirac's delta function, which leads to an instant impulse and then its effect vanishes. Here the Heaviside $\Theta$ function is present, meaning that even though at times $t<t'+|\mathbf{r}-\mathbf{r_s'}|/\upsilon_s$ the contribution is zero, the pulse emitted exactly at $t=t'$ has an everlasting effect, with the denominator playing the role of an attenuation factor, eventually disappearing at large values of $t$. The full wave describing this effect can be constructed as a superposition of wave modes with velocity values ranging from zero to $\upsilon_s$ (velocity of the outermost wave - the so-called Huygens surface). This is known as the "tail" of the Green's function leading to a non-sharp wave propagation, in contrast to the (3+1)-dimensional case \cite{dai_origin_2013}.

Using the two-dimensional Green's function we get
\begin{align*}
\phi_j&=-\frac{\mu}{\pi}\int d^2\mathbf{r'}\int dt'\frac{\Theta(\upsilon_s(t-t')-|\mathbf{r}-\mathbf{r'}|)}{\sqrt{(t-t')^2-|\mathbf{r}-\mathbf{r'}|/ \upsilon_s^2}}\rho_j,\\
 A_{ij}&=-\frac{1}{2\pi}\int d^2\mathbf{r'}\int dt'\frac{\Theta(\upsilon_s(t-t')-|\mathbf{r}-\mathbf{r'}|)}{\sqrt{(t-t')^2-|\mathbf{r}-\mathbf{r'}|/ \upsilon_s^2}}J_{ij}  ,
\end{align*}
where $\upsilon_s$ is the propagation velocity equal to $\sqrt{2\mu}$ for both fields, and $t'$ and $\mathbf{r'}$ are the retarded coordinates. By using the definitions of the density and current from the previous section and performing the integration in space, we arrive at:
\begin{equation}
 \phi_j=-\frac{\mu \epsilon_{ji}b_i}{\pi}\int \frac{\Theta(\upsilon_s(t-t')-|\mathbf{r}-\mathbf{r'_s}|)}{\sqrt{(t-t')^2-|\mathbf{r}-\mathbf{r'_s}|/ \upsilon_s^2}}dt'
 \vspace{2mm}
\end{equation}
\begin{equation}
A_{ij}=-\frac{\epsilon_{\left(ik\right.}\upsilon_{\left.j\right)}b_k}{2\pi}\int \frac{\Theta(\upsilon_s(t-t')-|\mathbf{r}-\mathbf{r'_s}|)}{\sqrt{(t-t')^2-|\mathbf{r}-\mathbf{r'_s}|/ \upsilon_s^2}}dt'
\vspace{2mm}
\end{equation}
The above integrals have considerable complexity and only in
some simple cases yield results of elementary functions in a closed form. We will thus be using the formalism of near-field approximation adopted by Lazar \cite{lazar_nearfield}, where the lower time limit solution is that of regular elastostatics with the defect immobile until $t=0$. The lower limit of the above integral is switched from $-\infty$ to zero and the upper one has the relativistic boundary value of $t-|\mathbf{r}-\mathbf{r'}|/\upsilon_s$. By assuming that the defect does not move too far from the observational point, we can take $|\mathbf{r}-\mathbf{r'}|\approx d$. The final result is
\begin{equation}
\phi_j=\frac{\mu \epsilon_{ji}b_i}{\pi}\ln\left(\frac{d}{\upsilon_st+\sqrt{\upsilon_s^2t^2-d^2}}\right)\Theta(\upsilon_st-d),
\end{equation}
\begin{equation}
A_{ij}=\frac{\epsilon_{\left(ik\right.}\upsilon_{\left.j\right)}b_k}{2\pi}\ln\left(\frac{d}{\upsilon_st+\sqrt{\upsilon_s^2t^2-d^2}}\right)\Theta(\upsilon_st-d)  .
\end{equation}
As it can be seen, just like the regular charged particle case a connection between the field potentials is given in the form of $A \propto \upsilon^{-1}\cdot \phi$. This time the usual relation is inverted because we worked with $\Tilde{C}^{-1}_{ijkl}$ in our equations of motion.

By making use of Eq. \eqref{gaugefields}, we find the elastic analog of Jefimenko's equations for a dislocation
\begin{widetext}
\begin{equation}
E^{\sigma}_{ij}=\frac{[-2\mu \epsilon_{ik}b_k n_j+\epsilon_{\left(ik\right.}\upsilon_{\left.j\right)}b_k(n_m\upsilon_m)]}{2\pi d}\left(\frac{A(t,d)}{1-n_i\beta_i}\right)+\frac{\epsilon_{\left(ik\right.}\upsilon_{\left.j\right)}b_k}{2\pi}\frac{1}{B(t,d)}\left(\upsilon_s+\frac{\upsilon_s^2t}{B(t,d)-\upsilon_st}\right)-\frac{\epsilon_{\left(ik\right.}\dot{\upsilon}_{\left.j\right)}b_k}{2\pi}\ln\left(\frac{d}{B(t,d)}\right),
\label{Electric}
\end{equation}
\begin{equation}
B_i=\left(\frac{b_{\left[k\right.}n_{\left.i\right]}\upsilon_k+(b_jn_j)\upsilon_i}{2\pi d}\right)\left(\frac{A(t,d)}{1-n_i\beta_i}\right),
\end{equation}
\end{widetext}
where in the above $n_i$ is the unitary vector component of $\mathbf{r}-\mathbf{r_s}$ and we have also introduced the relativistic factors $\beta_i=\upsilon_i/\upsilon_s$, $A(t,d)=1+d^2/[B(t,d)(B(t,d)-\upsilon_st)]$ and $B(t,d)=\upsilon_st+\sqrt{\upsilon_s^2t^2-d^2}$.\\
We can now invert the duality in order to find the dynamical time dependent stress-tensor components to be
\begin{equation*}
\sigma_{xx}=\frac{[2\mu b_xn_y-\upsilon_yb_x(n_i\upsilon_i)]}{2\pi d}\left[\frac{A(t,d)}{1-n_i\beta_i}\right]-\hspace{10cm}
\end{equation*}
\begin{equation*}
-\frac{\upsilon_yb_x}{2\pi}\frac{1}{B(t,d)}\left(\upsilon_s+\frac{\upsilon_s^2t}{B(t,d)+\upsilon_st}\right)+\frac{\dot{\upsilon}_yb_x}{2\pi}\ln\left(\frac{d}{B(t,d)}\right),
\end{equation*}
\begin{equation*}
\sigma_{yy}=\frac{[-2\mu b_yn_x+\upsilon_xb_y(n_i\upsilon_i)]}{2\pi d}\left[\frac{A(t,d)}{1-n_i\beta_i}\right]+\hspace{10cm}
\end{equation*}
\begin{equation*}
+\frac{\upsilon_xb_y}{2\pi}\frac{1}{B(t,d)}\left(\upsilon_s+\frac{\upsilon_s^2t}{B(t,d)-\upsilon_st}\right)-\frac{\dot{\upsilon}_xb_y}{2\pi}\ln\left(\frac{d}{B(t,d)}\right),
\end{equation*}
\begin{equation*}
\sigma_{xy}=\frac{[-4\mu b_xn_x+(\upsilon_xb_x-\upsilon_yb_y)(n_i\upsilon_i)]}{4\pi d}\left[\frac{A(t,d)}{1-n_i\beta_i}\right]+   \end{equation*}
\begin{equation*}
+\frac{(\upsilon_xb_x-\upsilon_yb_y)}{4\pi}\frac{1}{B(t,d)}\left(\upsilon_s+\frac{\upsilon_s^2t}{B(t,d)-\upsilon_st}\right)-
\end{equation*}
\begin{equation*}
\hspace{4cm}-\frac{(\dot{\upsilon}_xb_x-\dot{\upsilon}_yb_y)}{4\pi}\ln\left(\frac{d}{B(t,d)}\right).
\end{equation*}
From the static limit of the theory and by requiring a symmetric stress tensor, another condition extracted is $b_xn_x=-b_yn_y$. Both of these are a result of our initial symmetry of $\partial_i\phi_j=\partial_j\phi_i$, since the field $\phi_j$ is the gradient of the scalar fracton field $\phi$.\\ The strains now can be easily given by making use of $u_{ij}=C_{ijkl}^{-1}\sigma_{kl}$. We present the full expression in Appendix \ref{appc}. Note that the strains are obtained by reversing the duality and applying the inverse of the four-rank elasticity tensor to the stresses, leading to a dependency on both elastic moduli $\mu$ and $\kappa$. This will not be the case if we set $\mu =\kappa$ in the derivation of the wave equations \eqref{eq:wave1} and \eqref{eq:wave2}.  It is evident that as the elastic moduli increase, the contributions of dislocation velocities to stresses and strains diminish.\\
In the special case now where the number of vacancies is equal to zero an extra constraint involving the Burgers vector and the defect velocity is established since we require the trace of the defect current to vanish. This leads to the equation $J_{ii}=J_{xx}+J_{yy}=\upsilon_xb_y-\upsilon_yb_x=0$, that needs to be satisfied.

\subsection{Fluctuating number of vacancies present in the crystal}
We note that elastodynamics is described by a system of field equations representing the transmission of elastic waves. This can be easily appreciated in the language of displacements since taking the curl of Navier's equations one obtains two wave equations for dilatational and rotational disturbances \cite{eringen_elastodynamics_1975}. This is not so easy to implement in the stress formulation of elastodynamics, that parallels our gauge theory formulation, and in consequence the stress approach is less developed (see e.g. \cite{ostoja-starzewski_ignaczak_2019}).\\
The most general situation occurs when we allow for the vacancies to fluctuate. Under these circumstances, the trace contribution cannot be eliminated and necessitates the inclusion of an additional factor in the equations of motion. Let us postulate that the density of vacancies is a spatio-temporal function, i.e., $n_d(\mathbf{r'},t)$. By employing the condition for the quadrupole from the preceding section, we arrive at the following formulation:
\begin{equation*}
\phi_j=-\frac{1}{2\pi}\int d^2\mathbf{r'}\int dt'\frac{\Theta(\upsilon_s(t-t')-|\mathbf{r}-\mathbf{r'}|)}{\sqrt{(t-t')^2-|\mathbf{r}-\mathbf{r'}|/ \upsilon_s^2}}\rho_j^*    
\end{equation*}
\begin{equation*}
A_{ij}=-\frac{1}{2\pi}\int d^2\mathbf{r'}\int dt'\frac{\Theta(\upsilon_s(t-t')-|\mathbf{r}-\mathbf{r'}|)}{\sqrt{(t-t')^2-|\mathbf{r}-\mathbf{r'}|/ \upsilon_s^2}}J_{ij}^*    
\end{equation*}
where in the above $\rho_j^*=2\mu\rho_j+(\mu-\kappa)\partial_jn_d$ and $J_{ij}^*=J_{ij}+\delta_{ij}\dfrac{2\kappa(\mu-\kappa)}{(\mu+\kappa)}\partial_tn_d$ are the modified sources. Enforcing the form of the derivatives of a non-constant vacancy-interstitial density poses a challenge, akin to what we encountered with the delta functions for defect density and current. The related integrals also become more arduous to resolve analytically, especially considering the relativistic effects arising from the Green's function. Consequently, we opt to pursue a different approach. Our strategy to obtain exact wave solutions in the form of Li\'{e}nard-Wiechert potentials from Eqs. \eqref{wave-phi} and \eqref{wave-alpha} requires that, on top of fixing a gauge, we need to impose an additional constraint
\begin{equation} \label{eq:cosntraint}
\partial_tA_{ii}+C\partial_i\phi_i=0.
\end{equation}
Eq. \eqref{eq:cosntraint} constitutes a general condition of how this process happens since the trace, related to non-topological defects, is now a part of the solution. In order to have a better intuition we can rewrite \eqref{eq:cosntraint} using the definition of electric field \eqref{gaugefields}
\begin{equation} 
(1-C)\partial_i\phi_i =2\kappa n_d.
\end{equation}
We stipulate that the density of the vacancy-interstitial is linked to the derivative of one of the gauge fields associated with the electric tensor field. This is done in such a manner that their ratio retains its constancy, as previously illustrated. This condition represents a particular scenario where solutions can be analytically generated, while simultaneously preserving contributions from the trace-full segment of the differential equations. Indeed, verification can be made that setting $C=1$ reverts to the situation where $n_d=0$. In consequence $C$ parametrizes the configuration of non-topological defects in the theory. 

In the next step we calculate the distribution of fields. The equations of motion read
\begin{equation}
\Box \phi_j+C^{*}\rho_j=0, \qquad \upsilon_{\phi}=\sqrt{\frac{\mu\left[\kappa(1+C)+\mu(1-C)\right]}{\kappa}},
\end{equation}
\begin{equation}
\Box A_{ij}+\frac{(\kappa-\mu)}{4\mu\kappa}(1-C)\partial_t^2A_{kk}\delta_{ij}+J_{ij}=0, \qquad \upsilon_A=\sqrt{2\mu}.
\label{Trace}
\end{equation}
where in the above we have $C^*=\dfrac{4\mu\kappa}{\left[\kappa(1+C)+\mu(1-C)\right]}$. We identify two wave equations characterized by different velocities. One important consequence of the above equations is that the vector charge density is coupled to the bulk modulus, as opposed to the case without vacancies or interstitials. The dipoles in the theory are now allowed to move parallel and perpendicular to the Burgers vector.

The final solution for this case will be mixing contributions for two waves propagating with two different velocities and a time difference in the propagation startup, evident through the different $\Theta$ functions in the solutions. We have
\begin{equation}
\phi_j=\frac{C^*\epsilon_{ji}b_i}{2\pi}\ln\left(\frac{d}{\upsilon_{\phi}t+\sqrt{\upsilon_{\phi}^2t^2-d^2}}\right)\Theta(\upsilon_{\phi}t-d),
\end{equation}
\begin{equation}
A_{xy}=\frac{(\upsilon_xb_x-\upsilon_yb_y)}{4\pi}\ln\left(\frac{d}{\upsilon_At+\sqrt{\upsilon_A^2t^2-d^2}}\right)\Theta(\upsilon_At-d),
\end{equation}
The solutions for $\phi_i$ and the off-diagonal component of $A_{ij}$ remain consistent with their previous form. However, the distinctiveness arises when considering the components of the trace of the tensor potential. Now, its contribution must be incorporated into the analysis to accurately depict the dynamics. By evaluating the trace of Eq. \eqref{Trace}, we arrive at
\begin{equation}
\Box (A_{xx}+A_{yy})+J_{xx}+J_{yy}=0, \quad \upsilon_{\text{tr}}=\sqrt{\frac{2\kappa\mu}{\mu(1-C)+C\kappa}}.
\end{equation}
while for the difference of the two components the differential equation is
\begin{equation}
\Box (A_{xx}-A_{yy})+(J_{xx}-J_{yy})=0  \qquad \upsilon_A=\sqrt{2\mu}
\end{equation}
The final solution for the diagonal components then reads:
\begin{align}
\underset{i=j}{A_{ij}}&=\frac{\epsilon_{ik}\upsilon_{\left(j\right.}b_{\left.k\right)}}{4\pi}\ln\left[\frac{d^2}{(\upsilon_{tr}t+\sqrt{\upsilon_{tr}^2t^2-d^2})B(t,d)}\right]+ \nonumber \\
&\hspace{1.25cm}+\frac{\epsilon_{ik}\upsilon_{\left[j\right.}b_{\left.k\right]}}{4\pi}\ln\left(\frac{\upsilon_{tr}t+\sqrt{\upsilon_{tr}^2t^2-d^2}}{B(t,d)}\right) .
\end{align}
In this case the glide constraint does not hold any more and the dislocations exhibit climb motion in addition to the glide movement along the Burgers vector.

\section{Dynamical Peach-Koehler force}\label{Sec6}
In this section we focus on the interaction between dislocations (for a recent review of the developments in this subject in the conventional formulation of elasticity see \cite{lubarda_dislocation_2019}). We derive the formula describing the force between two dislocations for the traceless theory. An analogous computation for the fluctuating number of vacancies and interstitials case is straightforward. The Lorentz force exerted by one dipole on another reads:
\begin{equation}
F_j=-p_i(E_{ij}+\epsilon_{jk}\upsilon_k B_i)=\epsilon_{jk}b_i(\sigma_{ik}+\upsilon_k\pi_i).
\end{equation}
The first term represents the usual Peach-Koehler force, which will have contributions from both gauge fields, which account for relativistic corrections depending on time and the defect's velocity, along with the usual term that encapsulates the material's elastic properties. The second term is the Biot-Savart analog for the elastic force interaction and it purely depends on the velocity. The full force written as a function of the Burgers vectors and the velocity reads
\begin{widetext}
\begin{align} \label{eq:PKforce}
F_{j1\rightarrow 2}^{\text{PK}}&=\frac{-2\mu(b_ld_l)n_j-\left[(1/2)(\upsilon_kb_k)d_j+2(\upsilon_kd_k)b_j-\upsilon_j(b_kd_k)\right]n_b\upsilon_b+n_j(b_k\upsilon_k)(d_m\upsilon_m)}{2\pi d}\left(\frac{A(t,d)}{1-n_i\beta_i}\right)+\nonumber \\
&+\frac{(1/2)(\upsilon_kb_k)d_j+(\upsilon_kd_k)b_j-\upsilon_j(b_kd_k)}{2\pi}\frac{1}{B(t,d)}\left(\upsilon_s+\frac{\upsilon_s^2t}{B(t,d)-\upsilon_st}\right)+\\ 
&+\frac{(1/2)(\dot{\upsilon}_kb_k)d_j+(\dot{\upsilon}_kd_k)b_j-\dot{\upsilon}_j(b_kd_k)}{2\pi}\ln\left(\frac{d}{B(t,d)}\right). \nonumber
\end{align}
\end{widetext}
The force contains also a contribution from the acceleration (if present) of the defect just like in electrodynamics. The first term on the right hand side of Eq. \eqref{eq:PKforce} is the regular "electrostatic" component in the low velocity limit. This can be seen if the relativistic correction is equal to unity. It happens that when one interchanges $b_i\longleftrightarrow d_i$ and then fixes $n_i\rightarrow -n_i$, the result will be the force on the first dislocation produced by the second one, $F^{PK}_{2\rightarrow 1}$. By adding them both for the total force of the system, contributions from the first terms will cancel each other out, but this is not the case for the rest of terms containing corrections from velocities (i.e. the magnetic part). This parallels result in electrodynamics. We know that two moving charged particles exert forces that are not reciprocal. The explanation is that the difference in rate of change of the system's momentum is carried out as electromagnetic radiation. In our case the stress field carries the disturbance as a consequence of the elastic radiation from a moving dislocation.
\section{Discussion}\label{Sec7}

In this paper we show that a dual gauge theory formulation of elasticity simplifies the analysis of defect dynamics. Using this simplification, we extend existent results focused on edge dislocations. We construct solutions to the equations of tensor gauge theory, dual to Cauchy’s isotropic elasticity in a gauge, that is analogous to the Lorenz gauge in electrodynamics, for an arbitrarily moving dislocation. We use this to determine elastic stresses and strains due to the moving defect. We discuss separately two physical scenarios, when non-topological defects have a fixed number and when they are fluctuating. The former case corresponds to dislocations obeying the glide constraint, which in the static limit has been studied. We generalize this result to account for vacancies and interstitials. This leads to a new constraint that both topological and non-topological defects must satisfy, that we formulate as a relationship between gauge potentials. We interpret this result as a physical mechanism that forces dislocations to excite non-topological defects if they move perpendicularly to their Burgers vector. Finally, we use the duality to extend the formula of Peach and Koehler for the interactions between dislocations. We give an analytic formula that describes the dynamical Peach-Koehler force for two dislocations with different Burgers vectors or charges for the case ofa  fixed number of non-topological defects. 

The motion of defects in crystal lattices is essential to our understanding of material strength and plasticity.  Nevertheless, despite much effort dedicated to the subject, solving the defect dynamics in a conventional formulation of elasticity has led only to partial results. On top of that, isotropic Cauchy elasticity describes only the most basic crystalline solid. This renders those results unfeasible to be generalized to more complicated systems involving anisotropy, rotational degrees of freedom or incommensurate lattices. On the other hand, the simplicity of our approach allows one to extend results presented here to ask more detailed questions on crystals, such as radiation or problems related to fast-moving dislocations as well as understand defect dynamics in more general theories of elasticity that can be formulated in the language of gauge theories.

Finally we note that the problem of defects has recently reappeared in the context of active solids. Generalizing our results to the case of non-Hermitian elasticity could shed light on the dynamics of defects in active media. This will be useful in understanding recent experiments on active solids with odd elasticity.

\begin{acknowledgments}
  PS acknowledges support form the Polish National Science Centre (NCN) Sonata Bis grant 2019/34/E/ST3/00405. LT was supported in part by the Deutsche Forschungsgemeinschaft through the cluster of excellence ct.qmat (Exzellenzcluster 2147, Project 390858490).
\end{acknowledgments}

\onecolumngrid
\begin{appendices}
\section{Inversion of the elasticity tensor} \label{AppA}
We present here the supplementary theory regarding the inversion of the elasticity tensor. The most general case represented in the form of the three projector operators, is given by
\begin{equation}
C_{ijkl}=p_0 P_{ijkl}^0+p_1 P_{ijkl}^1+p_2 P_{ijkl}^2.
\end{equation}
We define
\begin{equation}
P_{ijkl}^0=\frac{1}{2}\delta_{ij}\delta_{kl},
\end{equation}
\begin{equation}
P_{ijkl}^1=\frac{1}{2}(\delta_{ik}\delta_{jl}-\delta_{il}\delta_{jk}),
\end{equation}
\begin{equation}
P_{ijkl}^2=\frac{1}{2}(\delta_{ik}\delta_{jl}+\delta_{il}\delta_{jk})-\frac{1}{2}\delta_{ij}\delta_{kl},
\end{equation}
where $p_0$, $p_1$ and $p_2$ are functions of the elastic parameters $\kappa$ and $\mu$. The projectors satisfy the closure identity $P_{ijkl}^0+P_{ijkl}^1+P_{ijkl}^2=\delta_{ij}\delta_{kl}$ and $P_{ijkl}^aP_{klmn}^b=P_{ijmn}^a$ if $a=b$ and zero otherwise. The inversion then is easily performed giving
\begin{equation}
C_{ijkl}^{-1}=\frac{1}{p_0}P_{ijkl}^0+\frac{1}{p_1}P_{ijkl}^1+\frac{1}{p_2}P_{ijkl}^2,
\end{equation}
For the case of an isotropic solid, we require the system to be invariant under the improper rotations group O(2) satisfying the relation $P_{ijkl}^1\sigma_{ij}=2\omega_{kl}=0$, where $\omega_{ij}$ is the angular velocity. The above tensor then is only partially invertible due to the lack of rotational degrees of freedom. In this case $p_1=0$ and the inverse is defined only in the
invertible subspace. Using the above formulas the elasticity tensor and its inverted counterpart for this case are then given by
\vspace{1mm}
\begin{equation}
C_{ijkl}=(\kappa-\mu)\delta_{ij}\delta_{kl}+\mu(\delta_{ik}\delta_{jl}+\delta_{il}\delta_{jk})   , 
\end{equation}
\begin{equation}
C_{ijkl}^{-1}=\left(\frac{1}{4\kappa}-\frac{1}{4\mu}\right)\delta_{ij}\delta_{kl}+\frac{1}{4\mu}(\delta_{ik}\delta_{jl}+\delta_{il}\delta_{jk}) .
\end{equation}
For the rotated inverted tensor now, we have
\begin{equation*}
\Tilde{C}_{ijkl}^{-1}=\epsilon_{ia}\epsilon_{jb}\epsilon_{kc}\epsilon_{ld}C_{abcd}^{-1}  
=(\delta_{ij}\delta_{ab}-\delta_{ib}\delta_{ja})(\delta_{kl}\delta_{cd}-\delta_{kd}\delta_{cl})C_{abcd}^{-1}=C_{ijkl}^{-1}.    
\end{equation*}
proving Eq.\eqref{elasticitytensor} used in deriving the equations of motion.

\section{Fracton Elastostatics} \label{AppB}
\subsection{Disclination}
In the case of no time dependence in the equations of motion, the differential equation for the fractonic vector field is
\vspace{1mm}
\begin{equation}
\nabla^2 \phi_j=-\left(\frac{4\mu\kappa}{\mu+\kappa}\right)\rho_j\Longrightarrow \nabla^4\phi=\left(\frac{4\mu\kappa}{\mu+\kappa}\right)\rho.
\end{equation}
The scalar field $\phi$ acts as an Airy stress function and the differential equation at hand is a non-homogeneous biharmonic equation in two dimensions, appearing often in linear elasticity and linearized fluid mechanics.
The solution is given as a special case of the Mitchell solution with no radial dependence, for $\rho=q\delta^2(r-r_s)$. Here $q$ represents the disclination charge. A simple and easy way of solving the above is by noticing that $\nabla^2\ln(r-r_s)=2\pi\delta^2(\mathbf{r}-\mathbf{r}_s')$. This gives the Poisson differential equation:
\begin{equation}
\nabla^2\phi=\frac{2q\mu\kappa}{\pi(\mu+\kappa)}\ln(\mathbf{r}-\mathbf{r}'_s)
\vspace{1mm}
\end{equation}
The solution is given by having the potential drop to zero at the boundary $R$ of the material
\begin{equation}
  \phi=\frac{q}{2\pi}\frac{\mu\kappa}{(\mu+\kappa)}(\mathbf{r}-\mathbf{r}'_s)^2\ln\left(\frac{|\mathbf{r}-\mathbf{r}'_s|}{R}\right) 
\end{equation}
where again the logarithmic behavior is expected. The vector field $\phi_i$ has components
\begin{equation}
\phi_i=\frac{q}{2\pi}\frac{\mu\kappa}{(\mu+\kappa)}|r_i-r'_{is}|\left[2\ln\left(\frac{|\mathbf{r}-\mathbf{r}'_{s}|}{R}\right)+1\right]  .
\end{equation}
Upon inversion of Eq.\eqref{rotated} one can extract the stress and electric tensors:
\begin{align*}
\sigma_{xx}&=-\frac{q}{2\pi}\frac{\mu\kappa}{(\mu+\kappa)}\left[2\ln\left(\frac{\sqrt{x^2+y^2}}{R}\right)+\frac{2y^2}{x^2+y^2}+1\right],\\
\sigma_{yy}&=-\frac{q}{2\pi}\frac{\mu\kappa}{(\mu+\kappa)}\left[2\ln\left(\frac{\sqrt{x^2+y^2}}{R}\right)+\frac{2x^2}{x^2+y^2}+1\right],\\
\sigma_{xy}&=\sigma_{yx}=-E_{xy}=-E_{yx}=\frac{q}{2\pi}\frac{\mu\kappa}{(\mu+\kappa)}\frac{2xy}{x^2+y^2},
\end{align*}
where, $\sigma_{xx}=E_{yy}$ and $\sigma_{yy}=E_{xx}$. The above equations are in full agreement with those in \cite{braverman_topological_2021}.
The strain tensor is given by making use of $u_{ij}=C_{ijkl}^{-1}\sigma_{kl}$, to get
\begin{align*}
u_{xx}&=-\frac{q}{4\pi(\mu+\kappa)}\left[2\mu \cdot \ln\left(\frac{\sqrt{x^2+y^2}}{R}\right)+3\mu+\kappa\left(\frac{y^2-x^2}{y^2+x^2}\right)\right],\\
u_{yy}&=-\frac{q}{4\pi(\mu+\kappa)}\left[2\mu \cdot \ln\left(\frac{\sqrt{x^2+y^2}}{R}\right)+3\mu+\kappa\left(\frac{x^2-y^2}{x^2+y^2}\right)\right],\\
u_{xy}&=u_{yx}=\frac{q}{2\pi}\frac{\kappa}{(\mu+\kappa)}\frac{xy}{x^2+y^2}
\end{align*}
\subsection{Dislocation}
For the case of a dislocation now, we assume a density of the form $\rho_i=\epsilon_{ij}b_j\delta^2(\mathbf{r}-\mathbf{r_s})$ and get a solution
\begin{equation}
\phi_i=\frac{2\epsilon_{ij}b_j\mu\kappa}{\pi(\mu+\kappa)}\ln\left[\frac{|\mathbf{r}-\mathbf{r_s}|}{L}\right].
\end{equation}
For the electric field and the stress tensor we calculate
\begin{align*}
\sigma_{xx}&=\frac{2b_x\mu\kappa}{\pi(\mu+\kappa)}\frac{(y-y_s)}{(x-x_s)^2+(y-y_s)^2},\\
\sigma_{yy}&=-\frac{2b_y\mu\kappa}{\pi(\mu+\kappa)}\frac{(x-x_s)}{(x-x_s)^2+(y-y_s)^2},\\
\sigma_{yx}&=-\frac{2b_x\mu\kappa}{\pi(\mu+\kappa)}\frac{(x-x_s)}{(x-x_s)^2+(y-y_s)^2},
\end{align*}
where $\sigma_{xx}=E_{yy}$, $\sigma_{yy}=E_{xx}$ and $\sigma_{yx}=\sigma_{xy}=-E_{xy}=-E_{yx}$. The requirement for a symmetric stress tensor dictates that \(\partial_i\phi_j = \partial_j\phi_i\), resulting in the relation \(b_x n_x = -b_y n_y\). This relation further guarantees that the equilibrium condition \(\partial_j \sigma_{ji} = 0\) is fulfilled. Finally we obtain the expressions for the strains
\begin{align*}
u_{xx}=&\frac{1}{2\pi(\mu+\kappa)}\frac{b_x(\mu+\kappa)(y-y_s)-b_y(\mu-\kappa)(x-x_s)}{(x-x_s)^2+(y-y_s)^2},\\
u_{yy}=&\frac{1}{2\pi(\mu+\kappa)}\frac{b_x(\mu-\kappa)(y-y_s)-b_y(\mu+\kappa)(x-x_s)}{(x-x_s)^2+(y-y_s)^2},\\
u_{xy}=&u_{yx}=-\frac{b_x\kappa}{\pi(\mu+\kappa)}\frac{(x-x_s)}{(x-x_s)^2+(y-y_s)^2}.
\end{align*}

\section{Strain} \label{appc}
The strain field extracted by using the relation $u_{ij}=C_{ijkl}^{-1}\sigma_{kl}$ reads
\begin{equation*}
u_{ij}=\left[\left(\frac{\kappa-\mu}{4\pi\kappa}\right)\left(b_kn_k+\frac{\epsilon_{km}\upsilon_kb_m(n_l\upsilon_l)}{2\mu}\right)\delta_{ij}-\frac{1}{2\pi}\left(b_in_j+\frac{1}{2\mu}\epsilon_{\left(ik\right.}\upsilon_{\left.j\right)}b_k(n_l\upsilon_l)\right)\right]\left[\frac{A(t,d)}{1-n_i\beta_i}\right]+\hspace{3.5cm} 
\end{equation*}
\begin{equation*}
+\frac{1}{4\pi\mu}\left[\left(\frac{(\kappa-\mu)\mu}{2\kappa}\right)\epsilon_{km}\upsilon_kb_m\delta_{ij}-\epsilon_{\left(ik\right.}\upsilon_{\left.j\right)}b_k\right]\frac{1}{B(t,d)}\left(\upsilon_s+\frac{\upsilon_s^2t}{B(t,d)-\upsilon_st}\right)+   
\end{equation*}
\begin{equation}
\hspace{7.8cm}+\frac{1}{4\pi\mu}\left[\left(\frac{(\kappa-\mu)\mu}{2\kappa}\right)\epsilon_{km}\dot{\upsilon}_kb_m\delta_{ij}+\epsilon_{\left(ik\right.}\dot{\upsilon}_{\left.j\right)}b_k\right]\ln\left(\frac{d}{B(t,d)}\right)    .
\end{equation}
\end{appendices}
\twocolumngrid
\bibliographystyle{utphys2}
\bibliography{LienardWiechert}

\end{document}